\documentclass[english,nocopyrightspace,acmowned=false]{sig-alternate}
\pdfoutput=1

\usepackage{algorithm}
\usepackage{babel}
\usepackage{booktabs}
\usepackage{cancel}
\usepackage[tableposition=top]{caption}
\usepackage{filecontents}
\usepackage[T1]{fontenc}
\usepackage[multiple]{footmisc}
\usepackage{gensymb}
\usepackage[unicode=true,
  bookmarks=true,
  bookmarksnumbered=false,
  bookmarksopen=false,
  breaklinks=true,
  pdfborder={0 0 0},
  backref=false,
  colorlinks=false]{hyperref}
\usepackage[latin9]{inputenc}
\usepackage{listings}
\usepackage{mathdots}
\usepackage{mathtools}
\usepackage{mhchem}
\usepackage{microtype}
\usepackage{paralist}
\usepackage{pgfplots}
\usepackage{stackrel}
\usepackage{stmaryrd}
\usepackage{tikz}
\usepackage{url}

\usetikzlibrary{backgrounds}
\usetikzlibrary{calc}
\usetikzlibrary{fit}
\usetikzlibrary{positioning}

\hypersetup{
  pdftitle={SWIM: Synthesizing What I Mean},
  pdfauthor={Mukund Raghothaman and Yi Wei and Youssef Hamadi},
  pdfsubject={Software Engineering: Coding, Tools and Techniques; Automatic
    Programming: Program Synthesis},
  pdfkeywords={Free form queries, code search, idiomatic snippet synthesis,
    structured call sequences}
}

\begin{document}

\global\long\def\toolname{\textsc{swim}}

\global\long\def\csharp{{C\#}}
\lstdefinestyle{csharp}{
  language=[Sharp]C,
  basicstyle=\ttfamily,
  morekeywords={var},
  showstringspaces=false
}
\global\long\def\csc#1{%
  \lstinline[style=csharp,
             mathescape=true,
             breaklines,
             keepspaces,
             literate={\-}{}{0\discretionary{-}{}{}}]@#1@}
\global\long\def\cscm#1{\text{\csc{#1}}}

\global\long\def\typespace{\textrm{\textit{Types}}}
\global\long\def\csstmt{\textrm{\textit{stmt}}}

\global\long\def\ch{\textrm{\textit{ch}}}
\global\long\def\chcreation{\textrm{\textit{creation}}}
\global\long\def\chusage{\textrm{\textit{usage}}}
\global\long\def\chaction{\textrm{\textit{action}}}
\global\long\def\chunknown{\textrm{\textit{unknown}}}
\global\long\def\chempty{\textrm{\textit{empty}}}

\global\long\def\prob#1{\operatorname{Pr}(#1)}
\global\long\def\condprob#1#2{\prob{#1 \mid #2}}

\global\long\def\vector#1{\mathbf{#1}}
\global\long\def\scsvec#1{\vector{v}_#1}
\global\long\def\similarity{\operatorname{similarity}}

\title{SWIM: Synthesizing What I Mean\thanks{This work was done when Mukund
  Raghothaman was an intern in Yi Wei and Youssef Hamadi's group at Microsoft
  Research, Cambridge, and will be presented at ICSE 2016. The authors
  gratefully thank Abhishek Udupa for suggesting the name \toolname{}.
}}
\subtitle{Code Search and Idiomatic Snippet Synthesis}
\numberofauthors{3}
\author{
  \alignauthor Mukund Raghothaman \\
  \affaddr{University of Pennsylvania}
  \email{rmukund@seas.upenn.edu}
	\and
  \alignauthor Yi Wei \\
  \affaddr{Microsoft Research, Cambridge} \\
  \email{jasonweiyi@gmail.com}
  \and
  \alignauthor Youssef Hamadi \\
  \affaddr{Laboratoire d'Informatique \\
    \'{E}cole Polytechnique, Palaiseau}
  \email{yhamadi75@gmail.com}
}

\maketitle


\begin{abstract}
  Modern programming frameworks come with large libraries, with diverse
  applications such as for matching regular expressions, parsing XML files and
  sending email. Programmers often use search engines such as Google and Bing to
  learn about existing APIs. In this paper, we describe \toolname{}, a tool
  which suggests code snippets given API-related natural language queries such
  as ``generate md5 hash code''.
  The query does not need to contain framework-specific trivia such as the type
  names or methods of interest.

  We translate user queries into the APIs of interest using clickthrough data
  from the Bing search engine. Then, based on patterns learned from open-source
  code repositories, we synthesize idiomatic code describing the use of these
  APIs. We introduce \emph{structured call sequences} to capture API-usage
  patterns. Structured call sequences are a generalized form of method call
  sequences, with \csc{if}-branches and \csc{while}-loops to represent
  conditional and repeated API usage patterns, and are simple to extract and
  amenable to synthesis.

  We evaluated \toolname{} with $30$ common \csharp{} API-related queries
  received by Bing. For $70\%$ of the queries, the first suggested snippet was a
  relevant solution, and a relevant solution was present in the top $10$ results
  for all benchmarked queries. The online portion of the workflow is also very
  responsive, at an average of $1.5$ seconds per snippet.
\end{abstract}

\section{Introduction}
\label{sec:intro}

%


Modern software engineering is reliant on large standard libraries, such as the
.NET Framework class library, the Java SDK, and the Android SDK. These libraries
provide a large variety of pre-implemented functionality, such as for matching
regular expressions, parsing XML files, sending email, and platform-specific
features such as accessing the GPS sensors and the phone camera. When faced with
an API-related task, most programmers rely on search engines such as Google and
Bing. They seek answers to two main questions:
\begin{inparaenum}[(\itshape a\upshape)]
  \item what APIs to use to solve their specific problem, and
  \item how to write the code involving those APIs.
\end{inparaenum}
Developers will often read a few returned web pages to see if their code must
follow certain programming idioms, or common usage practices. For example, good
practice dictates that files be closed after I/O is complete, and data may be
transmitted via a socket only after a connection is successfully established. In
many cases, developers search for API usage examples on online code repositories
such as GitHub and Bitbucket, or directly in their company's proprietary code
bases. This learning process is possible due to widely available information
related to programming. However, a developer still needs to read multiple web
pages, and many programs written by others to learn about these APIs and their
usage patterns.

This paper introduces \toolname{}, a tool to automate some of this discovery
process. \toolname{} is a code generator whose input is a natural language query
in English, such as ``match regular expression'' or ``read text file'', i.e. the
usual queries that a developer would enter in a search engine. In response,
\toolname{} outputs snippets of \csharp{} code, such as that shown in
figure~\ref{fig:ch:motivation:regex}, which hopefully implement the task
described in the query. The query and the synthesized code snippets are
API-related, meaning that they require the use of APIs in the solution. Given
that most programmers heavily use API libraries in their daily development
activities, this is an important class of queries submitted to search engines.

\begin{figure*}[ht]
  \centering
  \tikzstyle{block} = [draw, rectangle, minimum height=3em, minimum width=25ex]

\scalebox{0.95}{
  \begin{tikzpicture}
    \node (query) {User query};
    \node [block, below=2em of query] (apiLookup) {API lookup};
    \node [below=2em of apiLookup] (rankedAPI) {Ranked APIs};
    \node [block, below=2em of rankedAPI, align=center] (rootSCSSeln)
          {Structured call \\ sequence selection};
    \node [block, below=2em of rootSCSSeln, align=center] (codeSynth)
          {Code \\ synthesis};
    \node [below=2em of codeSynth, align=center] (solnSnippet)
          {Solution \\ snippets};

    \draw [->] (query) -- (apiLookup);
    \draw [->] (apiLookup) -- (rankedAPI);
    \draw [->] (rankedAPI) -- (rootSCSSeln);
    \draw [->] (rootSCSSeln) -- (codeSynth);
    \draw [->] (codeSynth) -- (solnSnippet);

    \node [left=15ex of apiLookup.west, anchor=east, align=center]
          (queryAPIModel) {Query to \\ API model};
    \node [left=10ex of queryAPIModel.west, anchor=east, align=center]
          (clickthrough) {Clickthrough \\ data};

    \draw [->] (clickthrough) -- (queryAPIModel);
    \draw [->] (queryAPIModel) -- (apiLookup);

    \node [left=15ex of rootSCSSeln.west, anchor=east, align=center]
          (scsIndex) {Structured call \\ sequence index};
    \node [block, left=10ex of scsIndex.west, anchor=east, align=center]
          (scsExtrn) {Structured call \\ sequence extraction};
    \node [left=10ex of scsExtrn.west, anchor=east, align=center]
          (github) {GitHub \\ code corpus};
    \node [left=15ex of codeSynth.west, anchor=east, align=center]
          (varname) {Variable name \\ model};

    \draw [->] (github) -- (scsExtrn);
    \draw [->] (scsExtrn) -- (scsIndex);
    \draw [->] (scsIndex) -- (rootSCSSeln);
    \draw [->] (github) |- (varname);
    \draw [->] (varname) -- (codeSynth);

    \begin{scope}[on background layer]
      \node [draw, dashed,
                fit={(clickthrough) (queryAPIModel)
                     (github) (scsExtrn) (scsIndex) (varname)
                     ($(clickthrough.north) + (0, 2em)$)
                     ($(scsIndex.east) + (5ex, 0)$)
                     ($(varname.south) + (0, -2em)$)
                     ($(github.west) + (-5ex, 0)$)}] (offlineBox) {};
      \node [anchor=north west] (offlineLabel) at (offlineBox.north west)
            {Offline analysis};
    \end{scope}
  \end{tikzpicture}
}
  \caption{Architecture of the \toolname{} tool.}
  \label{fig:intro:arch}
\end{figure*}
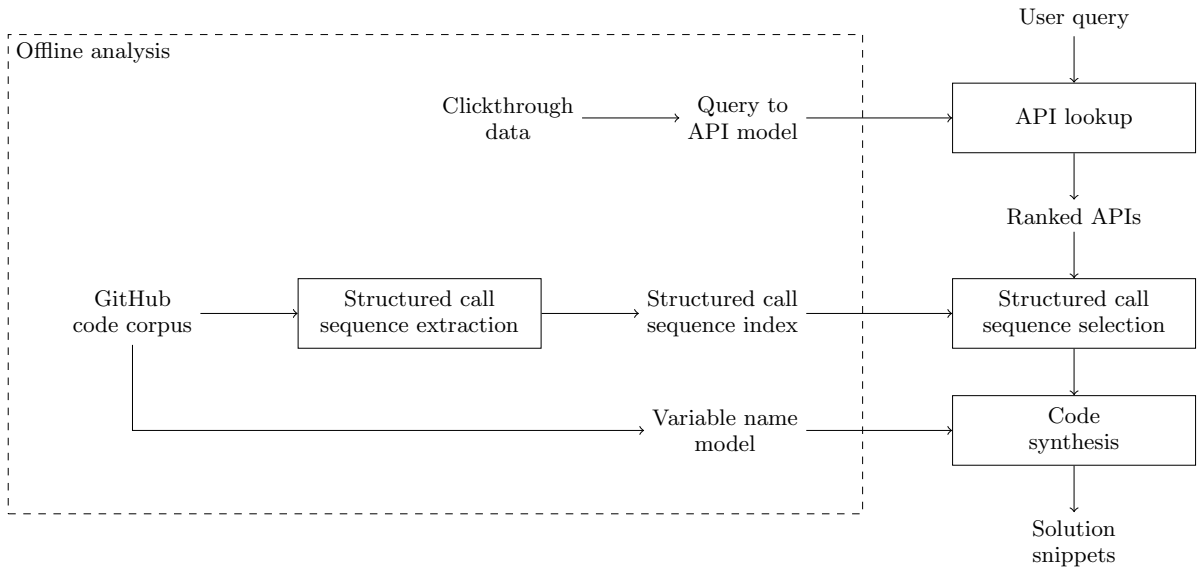

In this paper, the word API refers to a field or method from a class in the
framework. \toolname{} consists of two components: the first component, the
natural language to API mapper, suggests a set of APIs given a user query in
English. The second component, the synthesizer, generates code snippets using
the suggested APIs.

To suggest a set of APIs from a user query, the natural language to API mapper
builds a model of the form $\condprob{t}{Q}$, where $t$ is an API name, and $Q$
is a user query. This is the probability of the API $t$ appearing in a snippet
that solves the task described by $Q$. This model $\condprob{t}{Q}$ is learned
from clickthrough data collected by Bing. The clickthrough data is a log of
$(\textit{query}, \textit{URL})$ pairs which is recorded by most search engines.
Each $(\textit{query}, \textit{URL})$ pair indicates that the user clicked on
the result $\textit{URL}$ when presented with a list of results for
$\textit{query}$. For programming related queries, the returned web pages often
mention API names, or contain code snippets which invoke APIs. The clickthrough
data thus establishes a connection between English words in user queries and API
names in our target programming language, \csharp{}.

Note that the natural language to API name mapper only suggests which API names
should be used. There is often more detail to the use of an API than just the
method to be invoked or the field to be queried. This includes contracts such as
``\csc{T.foo()} may only be called after a call to \csc{T.bar()} returned
\csc{true}'', and best practices such as flushing a stream after writing to it.
This data and control flow information, along with other code artifacts such as
variable names, is not provided by the natural language to API mapper. This
is a deliberate design decision, because:
\begin{inparaenum}[(\itshape a\upshape)]
  \item we want to make use of most of the clickthrough data. Many clicked web
    pages may only mention the API name, without giving code snippets, so no
    program analysis can be performed. In such cases, we still want to record
    the fact that a particular API name is mentioned.
  \item The insight is that to solve a task, if a few key API names are given,
    the rest of the program is quite predictable, so a simpler model, which is
    consequently easier to implement and train, may suffice.
\end{inparaenum}

To synthesize code snippets from a set of suggested API names, the synthesizer
decides how to combine (some of) the APIs together to form a valid and
human-readable snippet.
There are several parts to code synthesis:
\begin{inparaenum}[(\itshape a\upshape)]
  \item deciding how the object is to be constructed,
  \item deciding the sequence of methods to be invoked, and fields to be queried
    and set in this object before the target API method may be invoked,
  \item the control flow between these object actions, and
  \item choosing appropriate variable names.
\end{inparaenum}
The synthesizer relies on another model, \emph{structured call sequences}, to
generate code with control- and data-flows. Structured call sequences describe
typical usage patterns for API classes. These usage patterns reveal how API
classes are used, for example, which method calls precede other method
invocations, and how control flows between these statements. Using structured
call sequences allows the synthesizer to generate code which covers the
suggested APIs from the natural language to API mapper, and at the same time,
obeys common coding conventions.

Structured call sequences are extracted from open-source projects on GitHub.
This is because for most classes $T$ in the API framework, GitHub contains many
more usage samples than can be extracted from web pages. For each supported
type, \toolname{} extracts structured call sequences from the source files in
the code corpus. In general, there may be multiple ways of using any given type,
and so multiple structured call sequences may be extracted for each type. They
are grouped by syntactic equality, and their occurrence frequencies are recorded
for later use in ranking the generated solutions.

Figure~\ref{fig:intro:arch} shows the overall architecture of \toolname{}. The
first thread builds the natural language to API mapper using clickthrough data
from Bing. The second thread analyses the GitHub code corpus and builds an index
of structured call sequences and a dictionary of variable names to be used
during code generation. Both threads are run offline. Finally, on receiving the
user query, we consult the natural language model to suggest a ranked list of
APIs, and find relevant structured call sequences from the index using those
APIs as search keys. These structured call sequences are synthesized into
solution snippets. This last thread is what is run online in response to a user
query.

To evaluate \toolname{}, we trained the natural language model with $15$ days of
clickthrough data from Bing, and learned structured call sequences from a corpus
of 25,000 open-source projects from GitHub. For each type, we extracted
structured call sequences from 10,000 source files using the type. We then asked
\toolname{} $30$ commonly occurring API-related queries from the Bing query
logs. A professional developer graded the results: for $70\%$ of the queries,
the first solution snippet was marked relevant, and for all the queries, a
relevant snippet was present in the top $10$ generated solutions. $88\%$ of the
chosen variable names were marked appropriate, and our response time was very
fast, averaging about $1.5$ seconds per produced snippet.

\paragraph{Contributions}
\label{par:intro:contrib}

In summary, the main contributions of this paper are:
\begin{enumerate}
  \item a technique to map natural language queries to API names;
  \item the concept of structured call sequences to express common API usage
    patterns, and an algorithm to extract structured call sequences from
    \csharp{} code;
  \item a synthesis algorithm to generate code snippets from structured call
    sequences; and
  \item a prototype implementation of these ideas in the tool \toolname{}, with
    experiments showing that relevant code snippets are generated for frequently
    asked API-related queries.
\end{enumerate}

\paragraph{Paper outline}
\label{par:intro:outline}

The rest of this paper is organized as follows. We introduce structured call
sequences, describe their extraction from the code corpus, and describe the
synthesis of code snippets in section~\ref{sec:ch}. Next, in
section~\ref{sec:nlp}, we introduce the natural language to API model, explain
how the model is trained from the search engine clickthrough data, and how user
queries are translated into structured call sequences for the synthesizer. We
then present a preliminary evaluation of \toolname{} in section~\ref{sec:eval},
and conclude with a summary of related work in section~\ref{sec:related}.

\section{Structured Call Sequences}
\label{sec:ch}

\subsection{Motivation}
\label{sub:ch:motivation}

In figure~\ref{fig:ch:motivation:regex}, we present \csharp{} code to match a
string against a regular expression. Focus on the object referred to by the
variable \csc{match}. It has type \csc{System.Text.Regular\-Expressions.Match},
and is created by the method \csc{Regex.Match(string)}, which accepts a single
argument \csc{input} of type \csc{string}. Next, the \csc{Groups} property of
the object is accessed depending on the value of \csc{Success}. Observe that
this pattern of object creation, method invocation and field accesses,
summarized as \csc{Regex.Match(string); if (get(Match.Success)) \{
get(Match.Groups) \}}, is a common way to use the \csc{Match} type: the
\csc{Match.Groups} field is only relevant if the input string matched the
regular expression, given by the field \csc{Match.Success}.

\begin{figure}
\begin{lstlisting}[style=csharp]
string pattern;
RegexOptions options;
var regex = new Regex(pattern, options);
string input;
var match = regex.Match(input);
if (match.Success)
{
  var groups = match.Groups;
}
\end{lstlisting}
  \caption{Example code to match a string against a regular expression.}
  \label{fig:ch:motivation:regex}
\end{figure}

As another example, in figure~\ref{fig:ch:motivation:readtextfile}, we present
code to read the contents of a text file using the
\csc{System.IO.Stream\-Reader} class. It is usual practice to release the
associated system resources by calling \csc{Stream\-Reader.close()} after I/O is
complete. In this case, the pattern of object usage may be summarized as:
\csc{new StreamReader(string); StreamReader.ReadToEnd(); StreamReader.Close()}.

\begin{figure}
\begin{lstlisting}[style=csharp]
string path;
var reader = new StreamReader(path);
reader.ReadToEnd();
reader.Close();
\end{lstlisting}
  \caption{Example code to read a text file using the \csc{Stream\-Reader}
    class.}
  \label{fig:ch:motivation:readtextfile}
\end{figure}

We introduce the term \emph{structured call sequences} to refer to these
patterns of object creation, method invocation, and field accesses. A structured
call sequence describes the object creation technique and a permissible sequence
of methods invoked and fields accessed, with control flow blocks such as
\csc{if} and \csc{while} to describe conditional and repeated usage patterns.
They can express more complicated object usage patterns than just the
construction technique or even specific sequences of invoked methods, but can
still be extracted easily, and can be readily synthesized into code snippets.
The thesis of this paper is that when grouped by syntactic equality, commonly
occurring structured call sequences largely describe idiomatic API usage.

We will formally define structured call sequences in
subsection~\ref{sub:ch:defn}, and then describe their extraction from the code
corpus in subsection~\ref{sub:ch:extraction}. Finally, in
subsection~\ref{sub:ch:synth}, we will describe how structured call sequences
are used in synthesizing code snippets to be presented to the user.

\subsection{Formal definition}
\label{sub:ch:defn}

\toolname{} works with a simple subset of the \csharp{} programming language. We
assume a finite set $\typespace$ of types. Each type $T$ has a finite set of
constructors, methods, and fields, and some methods and fields may be marked
\csc{static}. Methods are uniquely identified by their \emph{signature}: the
containing type, method name and the list of argument types. Each method
optionally has a return type, and is otherwise marked \csc{void}. Notably, the
language model disallows generic types, anonymous classes, first-class
functions, downcasts and exceptions.\footnote{This limitation is because the
synthesis a term with a given type is computationally hard in languages with
generics and first-class functions~\cite{SystemFTypeInhabitation}. Extending our
techniques to these language features is an important direction of future work.}


Given an object of type $T$, an individual program statement might either
\begin{inparaenum}[(\itshape a\upshape)]
  \item invoke a method \csc{$T$.method($a_1$, $a_2$, $\ldots$)}, with arguments
    $a_1$, $a_2$, \ldots, or
  \item get or set the value of a field \csc{$T$.field}.
\end{inparaenum}
Following the terminology of~\cite{Nguyen15-GraLan}, we term these atomic
constructs \emph{actions}:
\[
  \begin{array}{rcl}
    \chaction_T & \Coloneqq & \cscm{get($T$.field)} \mid \cscm{set($T$.field)} \\
                & \mid      & \cscm{$T$.method($a_1$, $a_2$, $\ldots$)}
  \end{array}
\]
If the member is \csc{static}, such as \csc{Console.WriteLine()}, then the
action can be performed without actually possessing an object of type $T$. If
the type of the queried field is $U$, or if the return type of the invoked
method is $U$, we write $\chaction_T : U$. As notational convenience, we also
treat constructors of the class $T$ as static methods named \csc{new} with
return type $T$.

A structured call sequence $\ch_T$ for a class $T$ begins with the creation of
an object with type $T$, and is a finite sequence of actions, together with
conditional statements and loops to represent repeated method invocations.
Formally, structured call sequences are productions of the following grammar:
\[
  \begin{array}{rcl}
    \ch_T & \Coloneqq & \chcreation_T \mid \chaction_T \mid \chunknown \\
        & \mid & \ch_{T1} ; \ch_{T2} ; \cdots ; \ch_{Tk} \\
        & \mid & \cscm{if ($\ch_{T1}$) \{\ $\ch_{T2}$ \} else \{\ $\ch_{T3}$ \}} \\
        & \mid & \cscm{while ($\ch_{T1}$) \{\ $\ch_{T2}$ \}} \\
    \chcreation_T & \Coloneqq & \chaction_U : T \\
  \end{array}
\]
The special construct $\chunknown$ indicates unknown object usage, for example,
when objects are passed to or returned by methods with unknown bodies. For ease
of presentation, we omit control structures such as \csc{for}-loops and \csc{do
while}-loops, even though these are also handled by \toolname{}.


\subsection{Extracting structured call sequences}
\label{sub:ch:extraction}

Our first problem is to scan the code corpus and extract, for each type $T$ in
the framework, all structured call sequences corresponding to $T$. We use the
recently developed Microsoft Roslyn compiler framework~\cite{Roslyn} to analyze
source files from the code corpus. Informally, Roslyn exports the compiler
services and associated parsing and analysis algorithms as a \csharp{} library.
It is a convenient framework because of two reasons:
\begin{inparaenum}[(\itshape a\upshape)]
  \item It gracefully handles errors in source files, and performs best-effort
    parsing, type resolution and variable binding in the presence of syntax
    errors and missing libraries. Because of its nature, we cannot expect all
    projects in the corpus to be free of compile-time errors, and diverse build
    systems make building or even identifying library dependencies infeasible.
    Furthermore,
  \item Roslyn transforms the source files into an AST representation with
    simple visitors, and this makes structured call sequence extraction
    straightforward.
\end{inparaenum}

The extraction algorithm works at the level of individual methods in the code
corpus. Let $v$ be a local variable of type $T$ which is not aliased by other
variables. We extract the structured call sequence $\ch_v$ describing its
lifetime by traversing the AST of the method body:
\begin{enumerate}
  \item For each assignment statement of the form \csc{$v$\ =\ $u$.member}, if
    we can resolve the referenced member to a (possibly static) member
    \csc{$U$.method()} of the framework, we produce the creation action
    \csc{$U$.method()}.
  \item For each method call \csc{$v$.method()} or field access \csc{$v$.field =
    \ $\ldots$} or \csc{var f =\ $v$.field}, we emit the corresponding action:
    \csc{set($T$.field)} or \csc{get($T$.field)} respectively.
  \item Given a sequence of statements $\csstmt_1; \csstmt_2; \cdots;
    \csstmt_k$, we extract the structured call sequence $\ch_i$ for the variable
    $v$ from each statement $\csstmt_i$, and produce the structured call
    sequence $\ch_1; \ch_2; \cdots; \ch_k$.
  \item For the conditional statement \csc{if ($\csstmt_1$) \{\ $\csstmt_2$ \}
    else \{\ $\csstmt_3$ \}}, we produce the structured call sequence \csc{if
    ($\ch_1$) \{\ $\ch_2$ \} else \{\ $\ch_3$ \}}, where $\ch_1$, $\ch_2$ and
    $\ch_3$ are the structured call sequences obtained from $\csstmt_1$,
    $\csstmt_2$ and $\csstmt_3$ respectively. \csc{While}-loops are similarly
    handled.
  \item Whenever $v$ is passed as an argument to another method, we insert a
    $\chunknown$ in $\ch_v$.
\end{enumerate}

Finally, we simplify the structured call sequences thus obtained by a few
straightforward rules. For example, \csc{$\ch_1$; if ($\chempty$) \{\ $\ch_2$ \}
else \{\ $\chempty$ \}}, where $\chempty$ is the empty sequence, is transformed
into \csc{$ch_1$; $\ch_2$}.

By not performing inter-procedural analysis or considering aliased variables,
our structured call sequence extraction technique is admittedly conservative. In
our anecdotal experience, while individual structured call sequences may be
ignored during extraction, because of the large number of source files used,
pervasive idioms are not missed. We postpone the development of more
sophisticated extraction techniques to future work.


\subsection{Synthesis from structured call sequences}
\label{sub:ch:synth}

We now consider the problem of transforming a linear call sequence $\ch_T$ into
a code snippet, given the user query and the chosen variable name $v$. We will
discuss the choice of variable names later in
subsection~\ref{ssub:ch:synth:var-names}, and section~\ref{sec:nlp} is devoted
to obtaining the structured call sequence $\ch_T$ from the user query.

The overall snippet synthesis procedure is described in
algorithm~\ref{alg:ch:synth:code-gen}. There are three important details,
corresponding to object creation, the synthesis of method arguments, and the
synthesis of boolean expressions, described in
subsections~\ref{ssub:ch:synth:creation},~\ref{ssub:ch:synth:arg}
and~\ref{ssub:ch:synth:boolean} respectively.

\begin{algorithm}[t]
  \caption{$\textit{code-gen}(\ch_T, v)$. Given the type $T$, linear call
    sequence $\ch_T$, and variable name $v$, the algorithm synthesizes the
    corresponding code snippet.}
  \label{alg:ch:synth:code-gen}

  \begin{enumerate}
    \item If $\ch_T = \chcreation_T$ or $\ch_T = \chaction_T$, the synthesis
      procedure is described in subsections~\ref{ssub:ch:synth:creation}
      and~\ref{ssub:ch:synth:arg} respectively.
    \item If $\ch_T = \ch_{T1}; \ch_{T2}; \cdots; \ch_{Tk}$ is a sequence, then
      the code snippet $\csstmt_i$ is synthesized for each element $\ch_{Ti}$.
      The concatenated snippet $\csstmt_1; \csstmt_2; \cdots; \csstmt_k$ is
      produced as output.
    \item If $\ch_T = \cscm{if ($\ch_{T1})$ \{\ $\ch_{T2}$ \} else \{\ $\ch_{T3}$
      \}}$, then a boolean expression \csc{boolExpr} is synthesized from
      $\ch_{T1}$ as described in subsection~\ref{ssub:ch:synth:boolean}. If
      $\ch_{T2}$ and $\ch_{T3}$ are synthesized into $\csstmt_2$ and $\csstmt_3$
      respectively, then we emit the snippet \csc{if (boolExpr) \{\ $\csstmt_2$
      \} else \{\ $\csstmt_3$ \}}. A similar procedure is used for
      \csc{while}-loops.
  \end{enumerate}
\end{algorithm}

\subsubsection{Object creation and tracer methods}
\label{ssub:ch:synth:creation}

Recall that a linear call sequence $\ch_T$ has a single creation action
$\chcreation_T$, which is also the first element in $\ch_T$. If $\chcreation_T$
is of the form \csc{$U$.method()}, where \csc{method()} is a static member of
the class $U$, then we can simply declare the variable $v$ as \csc{var\ $v$\ =
\ $U$.method($\ldots$)}. However, in the case that \csc{method()} is an instance
variable, we first need an object $u$ of type $U$. For example, constructing the
object \csc{match} in figure~\ref{fig:ch:motivation:regex} requires that we
already have an object of type \csc{Regex}.

Furthermore, it is conceivable that objects of type $U$ themselves maintain
state, and it is therefore insufficient to blindly construct objects of type $U$
and invoke \csc{method()}. Constructing the object $u$ is therefore similar to
the original snippet synthesis problem, except for the additional constraint
that it contain an invocation of \csc{method()}. We call \csc{method()} the
\emph{tracer} of interest.

We now describe the return value of $\textit{code-gen}(\ch_T, v)$, where $\ch_T
= \chcreation_T; \chusage_T$ and $\chcreation_T = \cscm{$U$.method()}$. Let
$\csstmt_T$ be the snippet synthesized by $\textit{code-gen}(\chusage_T, v)$.

Given the user query, let $\ch_U$ be the top-ranked structured call sequence
over $U$ which also contains an invocation of \csc{$U$.method()}, and $u$ be the
chosen name for the object of type $U$. Then $\textit{code-gen}(\chcreation_T;
\chusage_T, v)$ returns the output of $\textit{code-gen-tracer}$. The procedure
$\textit{code-gen-tracer}$ is similar to $\textit{code-gen}$, except that it
inserts the snippet $\csstmt_T$ immediately after the first invocation of
\csc{$U$.method()}.

For example, for the snippet in figure~\ref{fig:ch:motivation:regex}, we start
with the structured call sequence \csc{Regex.Match(string); if
(get(Match.Success)) \{ get(Match.Groups) \}}. We first synthesize the snippet:
\begin{lstlisting}[style=csharp, mathescape]
if (match.Success)
{
  var groups = match.Groups;
}
\end{lstlisting}
To construct the \csc{Regex} object, we then pick the structured call sequence
\csc{new Regex(string, RegexOptions); Regex.Match(string)} and merge the two
snippets to produce the final synthesized code.

While it is possible that the recursive object construction procedure may not
terminate, we have not observed this problem in practice. If necessary, we can
force termination after a pre-determined recursive depth with the default
snippet: \csc{default($U$).method()}.


\subsubsection{Synthesizing method arguments}
\label{ssub:ch:synth:arg}

A second difficulty is in synthesizing method arguments. Given a method
\csc{$T$.method($v_1$, $v_2$, $\ldots$, $v_k$)}, \toolname{} currently chooses
the default value for each argument (\csc{null} for reference types and zero for
value types):
\begin{lstlisting}[style=csharp, mathescape]
var $v_1$ = default($T_1$);
var $v_2$ = default($T_2$);
$\ldots$
var $v_k$ = default($T_k$);
$T$.method($v_1$, $v_2$, $\ldots$, $v_k$);
\end{lstlisting}
where $T_1$, $T_2$, \ldots, $T_k$ are the types of the respective arguments.
More involved schemes can be used to generate argument values, but would require
much greater computational resources, and are therefore not used.

\subsubsection{Boolean conditions}
\label{ssub:ch:synth:boolean}

The final interesting detail is in the synthesis of boolean expressions for
conditional statements and loop bodies. Synthesizing meaningful conditions would
require deep semantic knowledge: consider for example, distinguishing between
the pair of code snippets in figures~\ref{fig:ch:synth:condition-good}
and~\ref{fig:ch:synth:condition-bad}. Deciding which of these code snippets is
more ``standard'' requires understanding the semantics of the
\csc{IEnumerator.MoveNext()} method, and that a return value of \csc{true}
indicates that the iterator was successfully advanced to the next position.


We instead use the following simple procedure to obtain condition expressions.
Recall that every conditional in a linear call sequence has a single action. We
can readily convert every non-\csc{void} method \csc{$T$.method()} into the
boolean expression \csc{$T$.method($\ldots$) == default($U$)}, and every field
access \csc{get($T$.field)} into the boolean expression \csc{$T$.field ==
default($U$)}, where $U$ is the return type of the method or the type of the
field respectively. While we generally generate non-standard code, accessing the
correct fields in the conditions is usually valuable guidance to the programmer.

\begin{figure}
\begin{lstlisting}[style=csharp]
List list;
var enumerator = list.GetEnumerator();
while (enumerator.MoveNext())
{
  var current = enumerator.Current;
}
\end{lstlisting}
  \caption{Idiomatic conversion of the return value to a boolean.}
  \label{fig:ch:synth:condition-good}
\end{figure}

\begin{figure}
\begin{lstlisting}[style=csharp]
List list;
var enumerator = list.GetEnumerator();
while (!enumerator.MoveNext())
{
  var current = enumerator.Current;
}
\end{lstlisting}
  \caption{Non-standard conversion of the return value to a boolean.}
  \label{fig:ch:synth:condition-bad}
\end{figure}

\subsubsection{Picking variable names}
\label{ssub:ch:synth:var-names}

Consider the solution snippet from figure~\ref{fig:ch:motivation:regex} where
the variables \csc{pattern} and \csc{input} were instead called \csc{var1} and
\csc{var2} respectively. This hypothetical solution snippet is clearly inferior
as it obscures the role of the variables \csc{pattern} and \csc{input}.
Therefore, an important part of a good solution snippet is the choice of
descriptive variable names.

A similar problem has been considered by Raychev et
al~\cite{Raychev2015-BigCode}, in the context of deobfuscating JavaScript
programs. However, the statically-typed setting, and the fact that we are
synthesizing target code rather than analyzing it means that simpler techniques
suffice in our setting.

At each step during synthesis, we maintain a set of forbidden identifiers $F$.
This includes identifier names which have already been used and the set of
reserved \csharp{} keywords. Whenever we declare a new variable, we accumulate a
list of candidate names $C$, sorted by preference, and pick the first name in
$C$ which does not appear in the forbidden set $F$. If all candidate names are
unviable, we use the following simple fallback naming convention: the variable
name is the first non-forbidden identifier in the infinite list \csc{var1},
\csc{var2}, \csc{var3}, \ldots.

We will now describe the procedure to assemble the list of candidate names $C$.
For each method \csc{T.m()} in the API framework, we scan the code corpus and
construct a list $l$ of name-frequency pairs, $(\textit{name}_1, n_1),
(\textit{name}_2, n_2), \ldots$, where $n_i$ is the number of times the result
of the invocation \csc{T.m()} was assigned to a variable named
$\textit{name}_i$. Now consider an object \csc{v} described by the structured
call sequence \csc{var v = T.m(); $\ldots$}. For this object, we choose the
candidate names $C$ to be the list of names $\textit{name}_1, \textit{name}_2,
\ldots$ in $l$, arranged in descending order of frequency. A similar algorithm
is used to choose the candidate names for each field assignment \csc{T.f}.

Thus, for example, we observed that objects constructed using the
\csc{new Regex(string, RegexOptions)} constructor are most frequently assigned
to a variable named \csc{regex}, and objects returned by the
\csc{Regex.Match(string)} method are most frequently stored in a variable named
\csc{match}. Note that the construction of the name-frequency lists is performed
offline and the online variable naming algorithm simply chooses the first
non-forbidden name in this list.

For objects intended as method arguments, the candidate name list $C$ is the
singleton list $[ \textit{name} ]$, where $\textit{name}$ is the formal name of
the argument in the method declaration.

\section{Mapping user queries to structured call sequences}
\label{sec:nlp}

Structured call sequences represent empirically observed API usage idioms, but
do not directly tell us which high-level problems they solve. Since the
\toolname{} synthesizer accepts a natural language query as input, we first need
to find a mapping from natural language queries to the \csharp{} API. In this
section we describe how we use query expansion to model this mapping and
explains how we train the model from clickthrough data.

\subsection{Query expansion}
\label{sub:nlp:expansion}

The mapping from a natural language query to API names can be modeled as
$\condprob{t}{Q}$, the probability of a \csharp{} API $t$ appearing in the
solution snippet, given the user query $Q$. The higher the probability, the more
likely it is that $t$ appears in the code to solve the task described in $Q$.
The synthesizer applies this model to the query $Q$, finds the most likely APIs
that should appear in the synthesized snippet, and uses the structured call
sequences extracted from the code corpus to output the appropriate snippets.

The key idea is to view the computation of $\condprob{t}{Q}$ as a query
expansion problem. Query expansion is a commonly used technique in search
engines, where the user input is usually vague. Experience and research have
shown that adding one or more words to the queries can enhance the precision of
the search result. This process is called query expansion. Usual candidates for
word expansion include synonyms of the words appearing in the user queries. In
our case, we want to find the API names that are relevant to the user query,
i.e. expand the user query with API names.

People have proposed many ways to formulate the query expansion problem. In this
paper, we follow the method proposed by Gao et al.~\cite{gao2012towards}, which
uses \emph{clickthrough data} to find relevant words for expansion. When a user
types a query in a search engine, and the engine returns a list of results, the
user may click on one or more links. Search engines typically record a lot of
information about this click, but in the present paper, we only consider the set
of pairs $(\textit{query}, \textit{URL})$, indicating the url $\textit{URL}$
the user clicked on in response to the search term $\textit{query}$.

For a programming-related query, the clicked web page will possibly contain one
or more program fragments. To find candidate API names $t$ for the expansion, we
look at code fragments appearing on those web pages. We examine text contained
within HTML tags such as \lstinline[basicstyle=\ttfamily]{<pre>},
\lstinline[basicstyle=\ttfamily]{<code>} and
\lstinline[basicstyle=\ttfamily]{<p>}, which are likely to contain code
fragments. We then use the \csharp{} parser from Roslyn to parse the text (the
text has been preprocessed before parsing, to correct obvious syntax errors),
and determine whether it is a fragment of \csharp{} code. Finally, API names are
extracted from the parsed code fragments. Besides code fragments, we also
collect API names that are mentioned in the text.

Let $P$ be the list of API names in the code fragment, in their appearance
order. Then a single clickthrough pair $(\textit{query}, \textit{URL})$ can be
represented as a set of $(Q, P)$ pairs (because there may be multiple fragments
on a web page). 

The mapping from the user query $Q$ to an API name $t$, or the probability of
$t$ being the expansion term given $Q$, is given by:
\begin{alignat}{1}
  \condprob{t}{Q} & = \condprob{t}{q_1, q_2, \ldots, q_n} \nonumber \\
                  & = \sum_{i = 1}^n \condprob{t}{q_i} \cdot \condprob{q_i}{Q},
                      \label{eq:nlp:expansion}
\end{alignat}
where $Q = [q_1, q_2, \ldots, q_n]$ represents the user query containing the
words $q_1$, $q_2$, \ldots, $q_n$; $\condprob{t}{q_i}$ is the probability of the
API $t$ given a single query word $q_i$; $\condprob{q_i}{Q}$ is the unsmoothed
unigram probability of the query word $q_i$ in the query $Q$.

Equation~\ref{eq:nlp:expansion} decomposes the calculation of $\condprob{t}{Q}$
into the calculation of two simpler probabilities $\condprob{t}{q_i}$ and
$\condprob{q_i}{Q}$. The former quantifies the connection between an API and a
single word in the user query. The latter quantifies how likely the query term
$q_i$ appears in queries; it serves as normalization. We now describe how we
estimate these two probabilities from data.

\subsection{Estimating $\condprob{t}{q}$}
\label{sub:nlp:estimation-t}

$\condprob{t}{q}$ represents the probability of the API element $t$ appearing in
the solution snippet, given the occurrence of the word $q$ in the user query. It
establishes the connection from English words to \csharp{} API elements. We
estimate this model using clickthrough data, which also links user queries to
web pages containing API names.

As described above, clickthrough data contains $(Q, P)$ pairs, where $Q$ is the
user query and $P$ is the list of API names appearing in code fragments on the
clicked web page. Note however that this still does not relate individual query
words $q \in Q$ to API elements $t \in P$. To solve the problem, we use a
standard procedure for training statistical word alignment
models~\cite{brown1993mathematics} by applying an expectation maximization (EM)
algorithm. The EM algorithm first initializes $\condprob{t}{q}$ to random values
for each $t$ and $q$, and then iteratively updates the probabilities to maximize
the likelihood of generating the training data.

As an example, if the user query is ``match regular expression'', then $Q =
[\mbox{``match''}, \mbox{``regular''}, \mbox{``expression''}]$. If the clicked
web page contains the snippet shown in figure~\ref{fig:ch:motivation:regex},
then the API sequence is $P = [\cscm{Regex}, \cscm{Match}, \cscm{Success},
\cscm{Groups}]$. Given enough data, the EM algorithm will eventually assign
values to $\condprob{t}{q}$ such that
$\condprob{\cscm{Regex}}{\mbox{``regular''}} >
\condprob{\cscm{Groups}}{\mbox{``regular''}}$.

\subsection{Estimating $\condprob{q}{Q}$}
\label{sub:nlp:estimation-q}

This probability quantifies how likely the query term $q$ is to appear in a
query and is calculated as follows:
\begin{alignat}{1}
  \condprob{q}{Q} & = \frac{\alpha_{q}}{\sum_{q^\prime \in Q} \alpha_{q^\prime}},
\end{alignat}
where $\alpha_q$ is the appearance frequency of $q$ in all possible queries. To
estimate $\alpha_q$, we use the same clickthrough data, focussing on just the
queries:
\begin{alignat}{1}
  \alpha_q & = \frac{\mbox{\# of times } q \mbox{ occurs in query log}}
                    {\mbox{Total term count in query log}}.
\end{alignat}

\subsection{Retrieving structured call sequences from user queries}
\label{sub:nlp:retrieval}

Given a user query $Q$, the $\condprob{t}{Q}$ model offers a list of possible
API elements, ranked by their probabilities. Each API element $t$ may be a
member of a different type $T$ in the framework. However, to generate code, the
\toolname{} synthesizer needs to start with a single structured call sequence.
This section describes how we use document similarity to choose the structured
call sequence from a ranked list of API names.

Let $A = [a_1, a_2, \ldots, a_N]$ be the list of all API names that the system
supports, where $N$ is the number of APIs. Then a real-valued vector of length
$N$ with each element chosen from the range $[0, 1]$ can represent the weight of
each API. Note that conceptually, this vector is very long, its length is equal
to the number of API names that are supported in the system. For example, the
current implementation of \toolname{} includes 30,345 types in common .NET
libraries and over 500,000 methods from those types. The vector is sparse, most
of its elements are zeros (or very small probability values if we perform
smoothing while training the $\condprob{t}{Q}$ model).

From a ranked list of probabilities $\condprob{t}{Q}$, we create the \emph{query
vector} by setting the corresponding element to the values of those
probabilities. For example, if $\condprob{\cscm{Regex}}{\mbox{``regular''}} =
0.1$ and $\condprob{\cscm{Match}}{\mbox{``regular''}} = 0.05$, then we set the
elements corresponding to the APIs \csc{Regex} and \csc{Match} in the query
vector to $0.1$ and $0.05$ respectively.

We can also similarly represent each structured call sequence $\ch$ by a vector
$\scsvec{\ch}$ of length $N$: for each API $t$ that appears in $\ch$, we set the
corresponding element in $\scsvec{\ch}$ to $1$, and all other elements are set
to $0$. We call this the \emph{structured call sequence vector} of $\ch$. The
synthesizer maintains a repository of vectors for all structured call sequences
mined from the code corpus.


With query vectors and structured call sequence vectors defined, the synthesizer
uses cosine similarity among those vectors to find the most relevant ones. The
cosine similarity function is widely used in information retrieval. It is
defined by the following formula:
\begin{alignat}{1}
  \similarity(A, B) & = \frac{\sum_{i = 1}^N A_i \times B_i}
                             {\sqrt{\sum_{i = 1}^N A_i^2} \times \sqrt{\sum_{i = 1}^N B_i^2}},
\end{alignat}
where $A$ and $B$ are two vectors of length $N$. Given two documents, the higher
the similarity, the more relevant the documents are to each other. We use the
implementation provided by the open-source information retrieval package
Lucene~\cite{Lucene}, which compares the query vector against all structured
call sequence vectors and returns a ranked list of structured call sequences.
These structured call sequences are then fed to the synthesis algorithm of
section~\ref{sub:ch:synth}.

\section{Evaluation}
\label{sec:eval}


In this section, we present some initial results from the \toolname{}
synthesizer. The tool is currently implemented as a \csharp{} library: we are
working on the design of an intuitive interface, so that more comprehensive user
studies and measurements of programmer productivity can be performed.

\begin{table*}
  \scriptsize
  \centering \caption{Benchmark queries. The columns are described in
    section~\ref{sub:eval:results}.}
  \label{tbl:queries}


  \begin{tabular}{l l c r r r r}
    \toprule
    Query & Typical APIs & FRank & \%Top5 & \%Top10 & Var (\%) & Time (in s) \\
    \midrule
    append strings              & StringBuilder.Append, ToString & 1& 100 & 100 & 36(83) & 30 \\
    append text file            & File.AppendText, AppendAllText & 4& 40& 40 & 12(100) &6 \\
    binaryformatter             & BinaryFormatter.Serialize, Deserialize & 1& 60& 80 & 21(76) &10 \\
    connect to database         & SqlConnection.Open& 1& 100& 100 & 22(95) &13 \\
    convert int to string       & Convert.ToString& 3& 20& 30 & 11(100) &6 \\
    convert string to int       & Convert.ToInt32& 1& 80& 50 & 13(92) &6 \\
    copy file                   & File.Copy& 3& 60& 40 & 16(100) &5 \\
    create file                 & File.Create, WriteAllText& 3& 40& 30 & 13(100) &7 \\
    current time                & DateTime.Now & 1& 80& 70 & 13(85) &10 \\
    download file from url      & WebClient.DownloadFile& 1& 100& 80 & 27(89) &13 \\
    execute sql statement       & SqlCommand.ExecuteNonQuery, ExecuteReader& 2& 60& 50 & 22(73) &17 \\
    generate md5 hash code      & MD5.ComputeHash & 2& 60& 40 & 17(88) &8 \\
    get current directory       & Directory.GetCurrentDirectory & 2& 20& 10 & 1(100) &8 \\
    get files in folder         & Directory.GetFiles & 2& 40& 40 & 12(100) &8 \\
    launch process              & Process.Start, WaitForExit; ProcessStartInfo& 8& 0& 20 & 19(84) &29 \\
    load bitmap image           & Bitmap.FromImage, FromFile & 1& 80& 90 & 65(77) &28 \\
    load dll                    & Assembly.Load & 1& 100& 80 & 20(95) &7 \\
    match regular expression    & Regex.Match; Match.Success & 1& 80& 70 & 40(90)  &23 \\
    open file dialog            & OpenFileDialog.ShowDialog, FileName & 1& 100& 90 & 20(50) &23 \\
    parse datetime from string  & DateTime.Parse & 1& 80& 40 & 46(89) &33 \\
    parse xml                   & XmlTextReader.Create, Read & 1& 40& 20 & 44(77) &33 \\
    play sound                  & SoundPlayer.Play, PlaySync & 1& 80& 40 & 11(100) &13 \\
    random number               & Random.Next, NextBytes, NextDouble & 1& 100& 100 & 27(93) &11 \\
    read binary file            & File.OpenRead, Read, FileStream.Read & 1& 80& 40 & 55(87) &26 \\
    read text file              & File.ReadAllText, StreamReader.ReadLine & 1& 80& 60 & 14(100) &23 \\
    send mail                   & SmtpClient.Send, MailMessage.From, MailAddress& 1& 60& 60 & 28(86) &24 \\
    serialize xml               & XmlSerializer.Serialize & 1& 100& 80 & 41(95) &12 \\
    string split                & String.Split, Regex.Split & 1& 60& 40 & 28(82) &8 \\
    substring                   & STring.Substring & 1& 40& 30 & 37(54) &13 \\
    test file exists            & File.Exists & 1& 20& 10 & 25(92) &9 \\
    \midrule
    Average                     &             & 1.6  & 65 & 54 & 25(88) &15 \\
    \bottomrule
  \end{tabular}
\end{table*}

\subsection{Setup}
\label{sub:eval:setup}

\toolname{} needs large amount of \csharp{} code to extract structured call
sequences, and clickthrough data to train the natural language to API mapper. To
prepare the data, we downloaded 25,000 \csharp{} projects from GitHub. These
projects together contain about 3 million files.

We extracted structured call sequences for 30,345 common .NET types. For each
type, we located 10,000 \csharp{} files where that type appears and used those
files for extraction.

To train the natural language to API mapping model, we used 15 days of
clickthrough data from the Bing search engine. We filtered the data to only
focus on queries that contain the keyword ``\csharp{}''. The training is done
through a standard implementation of expectation maximization algorithm.

To evaluate the synthesis process, we selected 30 API-related queries from the
Bing search log. These queries are frequently asked, and they cover various API
usages, from simple to more involved. The column labeled Query in
table~\ref{tbl:queries} lists the chosen queries. The typical APIs column of the
table lists some APIs that are commonly used to implement tasks described in the
queries, as suggested by the NLP model of section~\ref{sec:nlp}. Note that the
listed APIs are not exhaustive, because the same task can be implemented by many
different APIs in different ways. Only the most likely APIs are listed in the
table. The full list of generated solutions is uploaded as ancillary material
along with this paper.

\subsection{Experiment results}
\label{sub:eval:results}

We asked a professional developer to grade the top 10 \toolname{} solutions for
each benchmark query. The snippets were marked relevant / irrelevant, indicating
whether the developer thought that it implements the task described in the
query. We also asked the developer to annotate all the chosen variable names as
appropriate or inappropriate. A variable name was annotated as appropriate if it
adequately conveyed the purpose of the variable.

\subsubsection{Snippet relevance}
\label{ssub:eval:results:relevance}

The FRank column of table~\ref{tbl:queries} reports the rank of the first
generated solution that is relevant to the query. This metric is important
because most users will scan through the results from top to bottom. For the
benchmark queries, in 70\% of the cases, the first generated snippet is
relevant. This shows the synthesizer is able to locate the correct APIs and
further choose the likely control flow structures to generate snippets. Also
observe that in all cases, at least one of the top 10 solutions was marked
relevant.

The \%Top5 and \%Top10 columns of table~\ref{tbl:queries} report the percentage
of relevant snippets in the top 5 and 10 generated solutions. Observe that the
user queries are vague, and there are usually many ways to implement a given
task using different APIs, and so there is no single correct solution to a
query. By exploring different APIs and different usage patterns, the synthesizer
generates variations of the same topic, so the users can browse through and
understand differences among them. These two metrics quantify how relevant a
list of suggestions are to a query. On average, 65\% of the synthesized snippets
from the top 5 generated solutions, and 54\% from the top 10 solutions are
observed to be relevant. This suggests that the overall list of presented
solutions is itself relevant to the user.

\subsubsection{Variable name choices}
\label{ssub:eval:results:var-names}

The proper choice of variable names is an important part of program
comprehensibility, and particularly so in program synthesis. The column marked
Var (\%) in table~\ref{tbl:queries} lists the number of variable names that the
synthesizer chose for the 10 most relevant snippets. The numbers outside the
parentheses are the number of variable names, and the numbers inside the
parentheses are the fraction of meaningful names as annotated by the
professional developer. The numbers reveal that in majority of cases, 88\% on
average, the synthesizer is able to find meaningful variable names. It also
shows that for very specific tasks, such as ``random number'', ``serialize
xml'', the chosen variable names are more likely to be meaningful; while for
more general tasks such as ``substring'', the variable names contain more noise.
This is because the synthesizer chooses the variable names according to their
appearance frequency in GitHub repositories. For specific tasks, the
distribution of variable names given by programmers are more focused on a small
range of names, while for general tasks the variable name distribution tends to
be more uniform.

\subsubsection{Synthesis time}
\label{ssub:eval:results:time}

Finally, responsiveness was an important requirement while creating \toolname{}.
The column marked Time in table~\ref{tbl:queries} shows the time required by the
synthesizer to generate the top 10 solutions. The experiments were run on a
desktop workstation with a 3.6GHz processor and 16 GB of RAM. Observe that we
require an average of 1.5 seconds to produce each solution snippet, and believe
that this is responsive enough for practical use. Also note that the current
prototype synthesizer is not optimized and contains many redundant calls to
Roslyn and the reflection APIs. Better engineering is likely to further reduce
the response time by a large fraction.

\subsection{Examples of synthesized snippets}
\label{sub:eval:examples}

We now provide concrete examples of synthesized snippets and discuss the
behavior of \toolname{}. We will also describe the limitations of the current
tool and ideas for future improvements.

\begin{figure}
\begin{lstlisting}[style=csharp]
var value = default(string); 
System.Convert.ToInt32(value);
\end{lstlisting}
  \caption{Incorrect solution snippet for query ``convert int to string''.}
  \label{fig:eval:inttostring}
\end{figure}

Figure~\ref{fig:eval:inttostring} shows the top snippet for the query ``convert
int to string''. This is an incorrect snippet because the snippet converts a
string to an integer by using the \csc{Convert.ToInt32()} method, instead of
converting an integer to a string. The third solution (not shown here) generated
by \toolname{} actually chooses the right method \csc{Convert.ToString()}. In
this case, the natural language to API mapper favors \csc{ToInt32()} since it
happens much more often. In future work, natural language processing techniques
such as pattern detection can be used to disambiguate APIs. For example, if the
query contains pattern $T_1$ to $T_2$, where $T_1$ and $T_2$ are types, we then
require the input and output type of the synthesized snippets to be $T_1$ and
$T_2$.

\begin{figure}
\begin{lstlisting}[style=csharp]
var dlg = new OpenFileDialog();
dlg.Title = null;
dlg.InitialDirectory = null;
dlg.Filter = null;
dlg.FilterIndex = 0;
if (dlg.ShowDialog())
{
  var fName = dlg.FileName;
}
\end{lstlisting}
  \caption{Snippet for query ``open file dialog''.}
  \label{fig:eval:openfiledialog}
\end{figure}

Figure~\ref{fig:eval:openfiledialog} shows the top snippet for the query ``open
file dialog''. The snippet first initializes a \csc{FileOpenDialog} object, and
then sets a few properties such as title of the dialog, initial directory of the
file explorer location when the dialog starts, the file pattern filters, and the
filter index. Then, the snippet shows the dialog and gets the user selected file
(if any) when the dialog is closed. Notice that all the properties are
initialized to the default value of the corresponding types. In \csharp{}, the
default value for \csc{string} is \csc{null}, and for \csc{int} is \csc{0}. This
is the default behavior of the synthesizer. While this is appropriate for
properties such as \csc{InitialDirectory}, it is incorrect for properties such
as \csc{Filter}. The \csc{FileOpenDialog.Filter} property expects strings in a
certain format. For example, a filter to select text files and all files looks
like \csc{"Text Files (.txt)|*.txt|All Files (*.*)|*.*"}. Such properties are
common in API libraries. Other examples include database connection strings,
which is needed in the query ``open database connection'' and the datetime
format strings, such as \csc{"yyyy$-$mm$-$dd"}, in the query ``parse datetime
from string''.

Ideally, for properties required to be in a certain format, the synthesizer
should provide some common patterns, instead of just generating the type-wise
default values. The difficulty is in automatically determining which properties
require formats. Potential solutions include:
\begin{inparaenum}[(\itshape a\upshape)]
  \item scanning the documentation of class properties to detect the mention of
    particular formats; and
  \item  scanning code repositories to find properties which are frequently
    assigned constants and use heuristics to decide if those constants have some
    structure.
\end{inparaenum}

\begin{figure}
\begin{lstlisting}[style=csharp]
var startInfo = new ProcessStartInfo();
startInfo.FileName = null;
var process = Process.Start(startInfo);
process.WaitForExit();
\end{lstlisting}
  \caption{Snippet for query ``launch process'' (complete).}
  \label{fig:eval:launch1}
\end{figure}

Figure~\ref{fig:eval:launch1} shows the 8th solution snippet for the query
``launch process''. This snippet first creates a \csc{Process\-Start\-Info}
object and sets the \csc{File\-Name} property to \csc{null} (actually, the user
will set the property to the proper file name to launch), and then uses the
static method \csc{Process.Start()} from the \csc{Process} class to start the
process. The return value of the \csc{Start()} method is an object of type
\csc{Process}. Calling \csc{WaitForExit()} on the object waits for the launched
process to finish. To come up with this snippet, the synthesizer chooses the
root type \csc{Process} to first generate the third and the fourth line. And
then, since the \csc{Start()} method requires an object of type
\csc{ProcessStartInfo}, the synthesizer finds a structured call sequence of type
\csc{ProcessStartInfo} to come up with the first and second line.

However, if the synthesizer starts with the type \csc{Process\-Start\-Info},
then the result will not be a complete snippet. Figure~\ref{fig:eval:launch2}
shows this case. This snippet is the top snippet for the query ``launch
process''. It only includes the statements to initialize a ProcessStartInfo
object, but misses the statements on the Process class to start and terminate
the process. Thus, the snippet is incomplete. The reason for the incompleteness
is that after the part for \csc{ProcessStartInfo} is generated, the synthesizer
stops because the generated statements for \csc{ProcessStartInfo} do not rely on
any other objects. However, the synthesizer does not know \csc{ProcessStartInfo}
alone does not fully implement the user query.

To solve this problem, future work will allow the synthesizer to focus on more
than one root types, by modeling the joint probability $\prob{T_1, T_2}$,
representing the probability of $T_1$ and $T_2$ appearing together. If two types
are more likely to appear together, the synthesizer will generate fragments for
both types and combine them together. To estimate such joint probability, we may
need to do inter-procedure analysis when extracting structured call sequences.

\begin{figure}
\begin{lstlisting}[style=csharp]
var startInfo = new ProcessStartInfo();
startInfo.FileName = null;
startInfo.CreateNoWindow = false;
startInfo.RedirectStandardOutput = false;
startInfo.RedirectStandardError = false;
startInfo.UseShellExecute = false;
\end{lstlisting}
  \caption{Snippet for query ``launch process'' (incomplete).}
  \label{fig:eval:launch2}
\end{figure}

\section{Related Work}
\label{sec:related}

There is a large body of work on programmer assistance and snippet synthesis
tools. With the wide availability of open source software, there is a growing
realization that existing code corpuses can be used in program analysis and code
synthesis (including contemporary initiatives such as the DARPA MUSE
program~\cite{DARPA-MUSE}). In this section, we summarize existing work and
contrast it with the present paper.

\paragraph{Snippet synthesis as type inhabitation}
\label{par:related:type-inhabitation}

Traditional IDE tools such as IntelliSense are derived from early systems such
as Project Marvel~\cite{Kaiser88-ProjectMarvel}. These tools typically provide
interactive feedback listing the methods and fields in the highlighted object,
and expressions of appropriate type available for use in the highlighted
context.


The Prospector tool~\cite{Mandelin05-Jungloid} considered the problem of
synthesizing ``jungloids'': snippets of code which construct an object of type
$T_{\textit{out}}$, given an input object of type $T_{\textit{in}}$. Prospector
works with a very simple type system, where the set of types is finite, and a
set of pre-determined functions convert objects of one type into another.
Unfortunately, in languages with richer type-systems, such as with generics and
first-class functions, type inhabitation is computationally intractable. More
recent work~\cite{Gvero11-Interactive, Kuncak13-CompleteCompletion} focusses on
developing practical heuristics and techniques to rank completions so that
short, natural code snippets are ranked higher than longer snippets of code.
Synthesis of partial expressions~\cite{Perelman12-TypeDirectedCompletion} has
also been considered as a way to generalize IntelliSense, where the tool
automatically suggests expressions with holes that consume one or more objects
with known types, and emit an object whose type is optionally known.

Lastly, tools such as CodeHint~\cite{Galenson14-CodeHint} are very interesting
because they perform type inhabitation at runtime. At a very high level,
CodeHint is a debugger plugin which can be queried for expressions of a given
type or whose values satisfy some assertion.

One major limitation of these techniques is that the developer is required to
have some prior knowledge of the API framework (such as the names of types).
Expressing queries in natural language allows developers who are new to the
development environment to easily find their way around.

\paragraph{Typestate aware code completion}
\label{par:related:typestate}

Another shortcoming of type-inhabitation-based code completion techniques is
their ignorance of object state, which is central to the imperative programming
method. The notion of typestate was first considered by Strom and
Yemini~\cite{Strom86-Typestate}. While the original proposal required syntactic
extensions for API designers to describe typestate, there have been efforts to
automatically learn typestate by program analysis~\cite{Alur05-JIST} and from
code corpuses~\cite{Gruska10-6000Projects, Monperrus13-Voting}.

Efforts to describe API usage by
$n$-grams~\cite{Allamanis:2013:MSC:2487085.2487127} and method call
sequences~\cite{Mishne12-Typestate} can be seen as typestate-aware code
synthesis. \textsc{slang}~\cite{Raychev2014-Completion} similarly analyses
method call sequences from a code corpus, and uses statistical techniques to
insert method calls at designated holes in the user program. There are several
challenges with these approaches, including:
\begin{inparaenum}[(\itshape a\upshape)]
  \item sensitivity to algorithm parameters, such as $n$, which are difficult to
    set accurately, and
  \item difficulty with API usage idioms which are inexpressible as finite state
    machines (such as library-provided stack data structures), and where the
    next permissible methods depend on previous return values (for example,
    where \csc{IEnumerator.Current} contains a meaningful value only if the last
    call to \csc{IEnumerator.MoveNext()} returned \csc{true}).
\end{inparaenum}
Furthermore, given two method call sequences $s_1$ and $s_2$ from different
files in the code corpus, ``merging'' these into a single suggested call
sequence $s$ is difficult. In our work, we do not try to merge structured call
sequences, and instead group them by syntactic equality and suggest multiple
solution snippets.

Groums~\cite{Nguyen09-GrouMiner} are similar to structured call sequences, but
while structured call sequences deal with the lifetime of a single object,
groums relate data flows between multiple objects. Groums were initially
proposed to perform anomaly detection, but more recently,
GraLan~\cite{Nguyen15-GraLan} has been proposed as a similar statistical model
for code completion. Because \toolname{} synthesizes snippets from scratch,
rather than attempting to fill holes in existing programs, the simpler model
offered by structured call sequences suffices for our purposes.

\paragraph{Answering free-form queries}
\label{par:related:nl-search}


A major component of our problem setting is the use of free-form natural
language queries, while most existing work on snippet synthesis requires prior
knowledge of relevant types such as \csc{ProcessStartInfo} or
\csc{XmlTextReader}. The SNIFF system~\cite{Chatterjee09-SNIFF} attempts to
solve the same problem as us, but differs in technical details. In SNIFF, each
method call in the codebase is annotated with the text of the associated API
documentation while indexing. On receiving an input query, all annotated source
files matching the query are retrieved, and a ``type-based intersection'' of
these is returned as the synthesized code. The main differences from this system
are two-fold. First, the use of search engine clickthrough data rather than
relying on documentation text allows us to use a larger body of text and more
reliably convert natural language queries into the APIs of interest. Second,
because structured call sequences are extracted offline, rather than by online
codebase analysis, we can respond quickly to input queries, currently at an
average of $1.5$ seconds per synthesized snippet, even in our unoptimized
implementation.

In~\cite{Keivanloo:2014:SWC:2568225.2568292}, Keivanloo et al propose a method
to spot code examples from free-form user queries. The idea is to first group
code fragments together according to their structural similarity using clone
detection, and then a set of associative keywords, such as identifier names, are
extracted from each group of code. These keywords are matched against the user
query to retrieve and rank the code. The method is similar to SNIFF, where the
code is represented by the documentation of the APIs that are used in it.

Gvero et al~\cite{gvero2014} developed the anyCode tool to synthesize snippet
expressions from free-form queries. Given a query, the tool is able to
synthesize an expression invoking a single API that implements the desired task.
anyCode locates which API method to use by string matching. To handle the
problem of API name and search query term mismatch, anyCode includes words with
similar meanings to API names by using WordNet. anyCode also uses parse tree
from a natural language processing toolkit to find relations among variable
names and constant expressions mentioned in the query to put them in the
synthesized expression. anyCode is similar to what we built in Bing Developer
Assistant~\cite{BDA}, in which we also use NLP parse trees to handle variable
generation in code synthesis. The main difference between anyCode and our
current work is that anyCode is only able to synthesize an expression;
\toolname{} can synthesize snippets with multiple statements and control flows.
To synthesize such snippets, we face a much larger search space than anyCode
does, hence a code model describing popular usage patterns is key to making the
tool practical.

Le et al~\cite{Le:2013:SSS:2462456.2464443} introduced the SmartSynth tool,
which synthesizes mobile applications from free-form user descriptions.
SmartSynth focuses on a predefined set of APIs to use, and builds a model to map
words in user queries to the set of APIs. It also uses dataflow analysis to find
missing statements to synthesize. SmartSynth can generate larger snippets than
\toolname{}, but a user needs to provide a longer description to the tool. Another
difference is that SmartSynth focuses a predefined set of APIs while our tool
handles all possible APIs in the open domain.

In~\cite{allamanis2015}, Allamanis et al developed a bimodal model to map
natural language queries to snippets. The work builds a separate model for each
query type and is able to synthesize snippets for variations of the type of
query. For example, for the query type ``create array'', the method can
synthesize snippets for different ways to create arrays, such as ``create a 2d
array'', ``make int array''. The model for each query type is built manually by
fining all possible ways that people might ask for a query type, and this manual
process is expensive. In contrast, the work presented in current paper is fully
automated, but cannot understand subtle differences in the phrasing of a query.

\section{Conclusion}
\label{sec:concl}

In this paper, we described \toolname{}, a tool to synthesize API-related code
snippets given natural language queries. We mined API usage patterns, in the
form of structured call sequences, from open-source \csharp{} projects, and used
clickthrough data from Bing to map queries to the types and methods of interest.
We believe that structured call sequences are a fundamental empirical artifact
of API design, and that they can be used in numerous applications such as code
anomaly detection.

There are several potential directions of future work. First, better NLP
techniques would help to distinguish between similar APIs, such as
\csc{Convert.To\-Int32()} and \csc{Convert.To\-String()}. Second, better
structured call sequence extraction algorithms and handling of language features
such as exceptions would expand the range of expressible API-usage idioms.
Finally, modeling joint probability distributions would help to solve the
incomplete snippet problem of figure~\ref{fig:eval:launch2}.

\bibliographystyle{abbrv}
\bibliography{references}

\end{document}